\documentstyle[11pt]{article}

\begin{document}

\markboth{Edmundo M. Monte} 
{Embedding Versus Immersion in General Relativity}

\title{EMBEDDING VERSUS IMMERSION IN GENERAL RELATIVITY}
\author{EDMUNDO M. MONTE\thanks{e-mail: edmundo@fisica.ufpb.br and edmundomonte@pq.cnpq.br}\\
Departamento de Fisica, Universidade \\Federal da Paraiba, 58059-970, Jo\~{a}o Pessoa, Paraiba, Brazil.}

\maketitle


\begin{abstract}
We briefly discuss the concepts of immersion and embedding of space-times in
higher-dimensional spaces. We revisit the classical work by Kasner in which
he constructs a model of immersion of the Schwarzschild exterior solution
into a six-dimensional pseudo-Euclidean manifold. We show that, from a
physical point of view, this model is not entirely satisfactory since the
causal structure of the immersed space-time is not preserved by the
immersion.
\end{abstract}


\section{Introduction}	

The first example of immersion of space-times in higher-dimensional spaces
came from Kasner, in 1921, when he explicitly exhibited a local
isometric immersion of the Schwarzschild exterior solution.\cite{Kasner} Interest in
higher-dimensional theories of space-time seems to be increasing these days.
Actually,  the ideas that underlie the modern developments were first put
forward by Kaluza and Klein, who considered the possibility that our
spacetime may have extra dimensions.\cite{Kaluza} Currently there is a huge
literature on the subject of extra-dimensional theories, immersion and
embedding of space-times, mainly motivated by the generalized Kaluza-Klein
theory, string theory and the recent braneworld scenario.\cite
{ADD,RS,Maeda,ME,Kerner} In all these theoretical developments the geometrical
concepts of immersion and embedding play a very important role.\cite
{embedding} Our main goal in this paper is to give a precise meaning of
these concepts and briefly discuss the immersion obtained by Kasner. We
prove that in this case the immersion does not preserve the causal character
of curves, thereby not being quite appropriate as a physical model.

\section{Immersions and Embeddings}

Let $(V_{N},g)$ and $(V_{D},\eta )$ be two differentiable manifolds $V_{N}$
and $V_{D},$ with dimensions $N$ and $D$, endowed with the metric tensors $g$
and $\eta $, respectively. A differentiable application $Y:(V_{N},g)%
\rightarrow (V_{D},\eta )$ is called a {\it local and isometric immersion 
}if: \\
$\forall p\in U\subset V_{N}$, $dY_{p}:T_{p}V_{N}\rightarrow
T_{Y(p)}V_{D}$ is injective, where $U$ is an open neighbourhood of $p,$
and $g(v,w)=\eta (dY_{p}(v),dY_{p}(w))$, $\forall v,w\in T_{p}V_{N}$. If,
in addition, $Y$ is a homeomorfism onto $Y(U)$, where $Y(U)$ has the
induced topology of $V_{D}$, then $Y$ is called a \textit{local and
isometric embedding.} When $V_{N}\subset V_{D}$ and\ the inclusion $%
Y:V_{N}\subset V_{D}\rightarrow V_{D}$ is an embedding, we say that $V_{N}$
is a \textit{submanifold} of $V_{D}$.\cite{Manfredo}

Let us now consider $(V_{4},g)$ as a pseudo-Riemannian manifold,
corresponding to a solution of Einstein's equations, locally and
isometrically embedded in another pseudo-Riemannian manifold $(V_{D},\eta )$.
In local coordinates $\{x^{i}\}$and $\{Y^{\mu }\}$ of $V_{N}$, $V_{D}$, the
isometric condition takes the form 
\begin{equation}
g_{ij}=\eta _{\mu \nu }Y_{,i}^{\mu }Y_{,j}^{\nu },  \label{isometric}
\end{equation}%
where $\ g_{ij}$ is the induced metric on $V_{4};\ Y_{,i}^{\mu }$ $=\frac{%
\partial Y^{\mu }}{\partial x^{i}}$and the indices now take value in the
range $\mu =1,\ldots ,D$; $i=1,...,4$.

From the above definitions we see that the essential distinction between
immersion and embedding is of topological character. In fact, any embedding
is an immersion, while the converse is not always true.\cite{MEtop}

\section{The Immersion of Schwarzschild Space-time}

The Schwarzschild exterior solution, which describes the space-time geometry
outside a spherical symmetric body, is given, in spherical coordinates $%
(t,r,\theta ,\phi )$ by 
\begin{equation}
ds^{2}=(1-2m/r)dt^{2}-(1-2m/r)^{-1}dr^{2}-r^{2}(d\theta ^{2}+sin^{2}\theta
d\phi ^{2})
\end{equation}%
where $m$ is the so-called geometric mass. Actually, one can regard the
above geometry as defining two different space-times:\cite{HE,Oneill}\\
a) The exterior Schwarzschild space-time $(V_{4},g)$, where 
$V_{4}=P_{I}^{2}\times S^{2}\;$; $\;\\
P_{I}^{2}=\{(t,r)\in I\!\!R^{2}|\;r>2m\}$, and \\
b) The Schwarzschild black hole $(B_{4},g)$, with\newline
$B_{4}=P_{II}^{2}\times S^{2}\;$, $\;P_{II}^{2}=\{(t,r)\in
I\!\!R^{2}|\;0<r<2m\}.\;$\\
In both cases, $S^{2}$ denotes the sphere of radius 
$r$ and metric induced from $(2)$.

Now define the space $(E,g)=([P_{I}^{2}\cup P_{II}^{2}]\times S^{2},g)$ and
consider an application $Y:(V_{4},g)\rightarrow (V_{6},\eta )$, where the
pseudo-Euclidean metric $\eta $ is given by the line element: \cite{Kasner,MEtop} 
\begin{equation}
ds^{2}=\;dY_{1}^{2}+dY_{2}^{2}-dY_{3}^{2}-dY_{4}^{2}-dY_{5}^{2}-dY_{6}^{2}.
\label{metricsixd}
\end{equation}%
It is not difficult to see that the application $Y$  defined by the
equations \newline

$\left\{ 
\begin{array}{l}
Y_{1}=(1-1/r)^{1/2}\mbox{cos}t \\ 
\vspace{1mm}Y_{2}=(1-1/r)^{1/2}\mbox{sin}t \\ 
\vspace{1mm}Y_{3}=f(r),\;\;(df/dr)^{2}=\frac{1+4r^{3}}{4r^{3}(r-1)} \\ 
\vspace{1mm}Y_{4}=r\mbox{sin}\theta \mbox{sin}\phi \\ 
\vspace{1mm}Y_{5}=r\mbox{sin}\theta \mbox{cos}\phi \\ 
\vspace{1mm}Y_{6}=r\mbox{cos}\theta \vspace{1mm},%
\end{array}
\right.$\\
represents a local isometric immersion of $(V_{4},g)$ into\ $(V_{6},\eta )$. 
For convenience we are taking $2m=1$ in the Schwarzschild metric. Indeed,
a simple calculation shows that\ (\ref{isometric}) holds and that $%
dY_{p}:T_{p}V_{4}\rightarrow T_{Y(p)}V_{6}$ is injective, i.e., 
\[
dim[dY_{p}(T_{p}(V_{4}))]=4,
\]%
or, equivalently, 
\[
det(minor\;of\;matrix[dY_{p}]_{6\times 4},\;4\;order)=msin^{2}\theta sin\phi
\neq 0,
\]%
for $0<\theta <\pi $ and $0<\phi <\pi $ or $\pi <\phi <2\pi $.

It is easy to see that this isometric immersion does not constitue an
isometric embedding. Indeed, to verify this fact just consider a curve 
\begin{eqnarray*}
\alpha  &:&(a,b)\subset R\longrightarrow (V_{4},g) \\
t &\rightarrow &\alpha (t)\subset (U\subset V_{4},g)
\end{eqnarray*}%
The image of $\alpha (t)$ through the application $Y$ is ${Y(\alpha
(t))=(Y_{1}(\alpha (t))},\ldots ,Y_{6}(\alpha (t)))$. Now, if we take $%
a<t_{1}<t_{2}<b$, since $Y_{1}$ and $Y_{2}$ are periodic functions of $t$,
it is clear that it is possible to have 
\begin{equation}
\alpha (t_{1})\neq \alpha (t_{2})\;\;{ and }\;\;Y_{\mu }(\alpha (t_{1}))=Y_{\mu
}(\alpha (t_{2})),\mu = 1,2.  \label{closed}
\end{equation}
In other words, there is a neighbourhood $U\subset V_{4}$, where $Y$ is not a
homeomorfism. In the particular case where the curve $\alpha
(t)$ is timelike it follows directly from (\ref%
{closed})\ and the isometric condition (\ref{isometric}) that ${Y(\alpha (t))%
}$ is a closed timelike curve in $Y(U) \subset (V_{6},\eta)$. Obviously,
from the standpoint of physics this state of affairs is not entirely
satisfactory as the causal structure of the space-time $(V_{4},g)$ is not
preserved by the immersion.

It is natural at this point to wonder whether an isometric embedding of the
Schwarzschild space-time into a six-dimensional has ever been found. 
The positive answer to this question was given in 1959, with the
work of Fronsdal, who found a regular embedding of the Schwarzschild
exterior solution into a six-dimensional  pseudo-Euclidean space by using
hyperbolic functions instead of the trigonometric functions used by Kasner.\cite{Fronsdal} 

\section{Comment}

Finally, we conclude that, even though any immersion can be considered
locally as an embedding, it is important to be aware of the existence
sometimes of pathological domains in the immersed space-time, as in the case
of Kasner's immersion of the Schwarzschild exterior solution. Therefore one
should be careful with the terminology. From the point of view of physics,
models that employs embeddings are much useful than those that employs
immersions, since the latter do not preserve the topology of the immersed
manifold.
\section*{Acknowledgments}

We would like to thank Professor Marcos Maia for useful discussions. Thanks
also go to FAPES-ES/CNPq (PRONEX) and  FAPESQ-PB/CNPq (PRONEX) for financial
support. We are indebted to the referees, whose comments helped us to
improve the manuscript.

\end{document}